\documentclass[fleqn,usenatbib,usedcolumn]{mnras}

\usepackage[T1]{fontenc}
\usepackage{ae,aecompl}

\usepackage{multirow}
\usepackage{caption}
\usepackage{dsfont}
\usepackage{xcolor}
\usepackage{arydshln}
\usepackage{commath}

\usepackage{graphicx}
\usepackage{bbold}

\usepackage{graphicx}	
\usepackage{amsmath}	
\usepackage{amssymb}	
\usepackage{csquotes}
\usepackage{anyfontsize}
\usepackage{physics}
\usepackage{newtxtext,newtxmath}
\usepackage[capitalise]{cleveref}
\crefformat{footnote}{#2\footnotemark[#1]#3}

\makeatletter
\newcommand\footnoteref[1]{\protected@xdef\@thefnmark{\ref{#1}}\@footnotemark}
\makeatother

\newcommand{\noop}[1]{}

\title[Baryonic effects and the $S_8$ tension]{Implications of feedback solutions to the $S_8$ tension for the baryon fractions of galaxy groups and clusters}

\author[Salcido et al.]{Jaime Salcido$^{\href{https://orcid.org/0000-0002-8918-5229}{\includegraphics[scale=0.04]{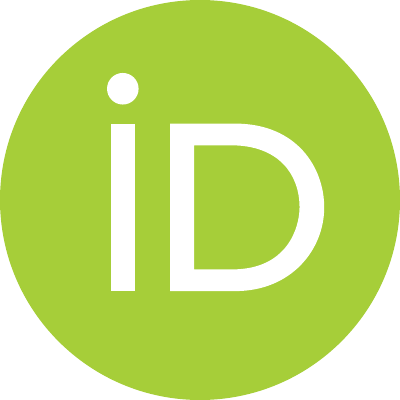}} 1}$\thanks{E-mail: \href{mailto:j.salcidonegrete@ljmu.ac.uk}{j.salcidonegrete@ljmu.ac.uk}} and
Ian G. McCarthy$^{\href{https://orcid.org/0000-0002-1286-483X}{\includegraphics[scale=0.04]{ORCIDiD.pdf}} 1}$\thanks{E-mail:
\href{mailto:i.g.mccarthy@ljmu.ac.uk}{i.g.mccarthy@ljmu.ac.uk}}
\\ \\
$^{1}$ Astrophysics Research Institute, Liverpool John Moores University, 146 Brownlow Hill, Liverpool L3 5RF, UK}

\date{Accepted XXX. Received YYY; in original form ZZZ}

\pubyear{2023}

\begin{document}
\label{firstpage}
\pagerange{\pageref{firstpage}--\pageref{lastpage}}
\maketitle

\begin{abstract}
Recent large-scale structure (LSS) surveys have revealed a persistent tension in the value of $S_8=\sigma_8\sqrt{(\Omega_{\mathrm{m}}/0.3)}$ compared to predictions from the standard cosmological model. Although some studies indicate that baryonic effects are too small to resolve this tension, others propose that more aggressive feedback mechanisms could reconcile differences between cosmic microwave background (CMB) measurements and low-redshift LSS observations. We investigate the role of baryonic effects in alleviating the $S_8$ tension. We extend the \texttt{SP(k)}\ model \citep{salcido_2023}, which was trained on hundreds of cosmological hydrodynamical simulations to map the suppression of the matter power spectrum to the baryon fraction in groups and clusters, to predict the required baryon fraction for a given $P(k)$ suppression. We then compare predictions from recent cosmic shear (weak lensing) analyses with the latest baryon budget measurements from X-ray and weak gravitational lensing studies. Our findings show that studies marginalising over baryonic effects while fixing cosmological parameters to a Planck-like cosmology predict strong $P(k)$ suppression and baryon fractions that are much lower than existing low-redshift baryon budget estimates of galaxy groups and clusters. Conversely, most studies that marginalise over both cosmological parameters and baryonic effects imply baryon fractions that are consistent with observations but lower values of $S_8$ than inferred from the CMB. Unless the observed baryon fractions are biased high by a factor of several, these results suggest that a mechanism beyond baryonic physics alone is required to modify or slow down the growth of structure in the universe in order to resolve the $S_8$ tension.
\end{abstract}




\section{Introduction}\label{sec:intro}

The $\Lambda$CDM cosmological model, deeply rooted in General Relativity, has long served as the foundation for our understanding of the Universe. It elegantly explains a multitude of cosmic observations, from the cosmic microwave background (CMB) to the large-scale structure (LSS) of matter at late times. However, recent large-scale structure surveys have delivered increasingly precise constraints on cosmological parameters, and they may be starting to reveal subtle cracks in the standard model \citep[see e.g.][]{Valentino_2021,Abdalla_2022}. 

One of these tensions is the so-called ``$S_8$ tension''. Weak gravitational lensing, a measure of the correlation in the distortions of the shapes of distant galaxies due to the intervening matter, has rapidly become an important test of the cosmological model. The quantity it best constrains is $S_8 = \sigma_8\sqrt{(\Omega_{\mathrm{m}}/0.3)}$, a combination of the present-day matter density $\Omega_{\mathrm{m}}$ and $\sigma_8$, the (linearly-evolved) amplitude of the matter power spectrum filtered on 8 Mpc/h scales. The best-fit value of the amplitude of the matter power spectrum appears to be in mild ($\approx 1.5-3\sigma$) tension with the predictions of the standard model fitted to the cosmic microwave background (see \citealt{Heymans2021,DES_Y3_2022} and references therein). This tension, though not statistically compelling on its own, has persisted for nearly a decade and spans several independent probes, each indicating tensions of similar significance and in the same direction.

There are three possible solutions to the ``cosmological tensions'' with the $\Lambda$CDM model. The first, and perhaps the most exciting possibility, is that the standard model of cosmology is incorrect and nature is governed by more ``exotic physics''. A wide range of beyond-$\Lambda$CDM models have been proposed to address the $S_8$ tension, including warm dark matter, decaying dark matter, and interacting dark energy (see, e.g., \citealt{Abdalla_2022} for a recent review). For instance, warm dark matter models can suppress small-scale structure formation \cite[e.g.][]{Viel_2013, Murgia_2017}, decaying dark matter can alter the growth rate of cosmic structures \cite[e.g.][]{Abell_2021,Elbers_2024}, and interacting dark energy models can modify the late-time expansion history and structure growth \cite[e.g.][]{Boehm_2005,Becker_2021}. While CMB measurements probe the early universe, LSS observations directly constrain the present-day cosmological parameters. Consequently, a deviation between these two parameter measurements might indicate that LSS has followed an evolutionary path distinct from what the standard model predicts. If this is correct, it could have significant consequences for fundamental physics.

The two remaining possibilities are that the measurements (or analysis thereof) of either the CMB data and/or LSS are flawed in some way. In terms of the CMB, it is unlikely that unknown systematics are responsible for the tension, at least in terms of the theory, as the thermal physics of the early Universe is well understood. While the measurements themselves could have systematic errors, independent measurements from previous large CMB missions, e.g., WMAP 9-year, have similar levels of tensions \citep[e.g.][]{Beutler_2016}.

In terms of the LSS analysis, current measurements are less precise than the CMB, and the modelling is more complicated. For instance, to obtain unbiased weak lensing cosmological constraints, accurate modelling of the non-linear matter distribution at scales of $0.1 \leq k \,\, [h/\mathrm{Mpc}] \leq 20$ is imperative (see e.g. \citealt{Huterer_2005} and \citealt{Hearin_2012}). This involves understanding not only the non-linear dark matter evolution due to gravity to the percent-level, but also the intricate feedback mechanisms associated with star formation and black hole growth that can significantly impact the distribution of matter on small scales. Failure to incorporate these effects can introduce biases in the inferred cosmological parameters from upcoming surveys like DESI, Euclid, and LSST \citep{semboloni_quantifying_2011,Semboloni_2013_2,chisari_modelling_2019,schneider_2020,castro_2021}. 

Recent cosmic shear surveys, such as the Kilo Degree Survey (KiDS), the Dark Energy Survey (DES) and the Hyper Suprime-Cam (HSC), have implemented different strategies to account for and quantify baryonic effects. The KiDS analysis incorporates these effects via a halo model framework, marginalising over a phenomenological `bloating' parameter that modulates the concentration of dark matter haloes \citep{Asgari2021, Heymans2021, Troster2021}. In contrast, DES initially addressed baryonic effects by applying scale cuts to eliminate small-scale data most susceptible to these influences \citep{derose_2019, Amon2022, Krause2021arXiv, Secco2022}. More recently, several studies have revisited the full DES dataset (without scale cuts) using baryonification techniques \citep{schneider_2015,arico_2021} to explicitly model the impact of baryonic physics \citep{arico_2023,chen_2023,Bigwood_2024}, while \cite{Terasawa_2024} use the halo model of \citep{mead_hydrodynamical_2020} to explore the baryonic effect signature in the HSC Year3 cosmic shear data \citep{L1_2022}. Additionally, a combined KiDS cosmic shear and Sunyaev-Zel’dovich effect analysis \citep{troster_2022} utilised the physically motivated halo model of \cite{mead_hydrodynamical_2020}, while a joint cosmic shear analysis involving DES, KiDS, and the HSC applied the \texttt{BACCOemu} emulator \citep{arico_2021} to explore baryonic effects \citep{Garcia_2024}. These studies have identified baryonic signatures at $\approx 2-3\sigma$ level, indicating that current observations are indeed sensitive to such effects.

In this paper, we investigate whether baryonic effects can alleviate the $S_8$ tension observed in cosmic shear (weak lensing) galaxy surveys. Previous studies based on cosmological hydrodynamical simulations suggest that the impact of baryonic effects may be too small to fully account for the observed tension \citep{mccarthy_bahamas_2018, McCarthy_2023}. Nevertheless, recent work by \citet{amon_2022}, \citet{preston_2023}, and the joint analysis of the DES cosmic shear with the kinematic Sunyaev-Zel’dovich (kSZ) data from the Atacama Cosmology Telescope \citep[ACT, ][]{ACT_2020} in \cite{Bigwood_2024}, have proposed that by considering more aggressive feedback mechanisms--beyond those typically modelled in simulations--it may be possible to reconcile the cosmological parameters derived from CMB measurements with those inferred from low-redshift LSS observations. 

To explore this hypothesis, we exploit the tight correlation between the suppression of the matter power spectrum and the baryon fraction within galaxy groups and clusters \citep{van_daalen_2020}. Specifically, we extend the \texttt{SP(k)} model introduced by \cite{salcido_2023}, which was calibrated using a comprehensive suite of 400 cosmological hydrodynamical simulations, to predict the baryon fraction required for a given level of $P(k)$ suppression. We then input suppression functions derived from various KiDS and DES analyses into the \texttt{SP(k)} model, translating these into predictions for baryon fractions. These predicted fractions are subsequently compared with the most recent baryon budget measurements, drawn from high-resolution X-ray observations in the XXL survey and weak-lensing mass measurements from HSC presented in \citet{Akino_2022}.

This methodology provides a stringent test of cosmological inferences from weak lensing surveys, as accurately modelling baryonic effects is crucial to avoid conflicts with other astronomical observations, such as the baryon fractions observed in galaxy groups and clusters. By evaluating whether the baryon fractions inferred from the suppression of the matter power spectrum are consistent with observed values, we aim to assess the extent to which baryonic effects may contribute to resolving the $S_8$ tension.

The present study is structured as follows. In Section \ref{sec:spk} we describe \texttt{SP(k)} empirical model that provides the mapping between the observable baryon fractions of groups/clusters and the suppression of the matter power spectrum, $P(k)$. In Section \ref{sec:ispk} we develop an \textit{`inverted'} version of the \texttt{SP(k)} model  to predict the required baryon fraction from a given $P(k)$ suppression, and test its accuracy against hydrodynamical simulations. In Section \ref{sec:res} we model the required baryon fractions from several recent weak lensing studies and test them against observations. Finally, in Section \ref{sec:Conclusions} we summarise our findings.

\begin{figure}
\centering 
\includegraphics[width=0.48\textwidth]{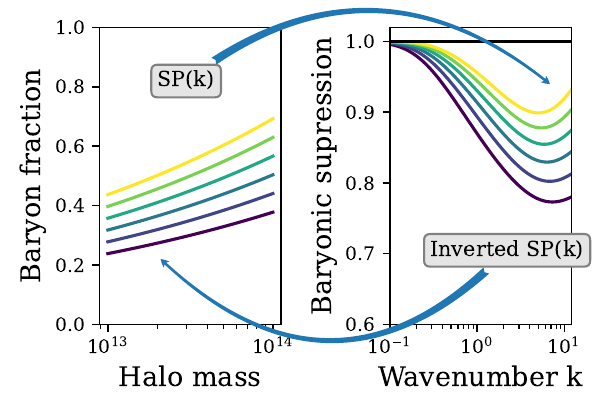}
 \vspace{-1.5em}
 \caption{Cartoon representation of the \texttt{SP(k)} model and its inverted form in \cref{eq:inv_spk}. \texttt{SP(k)} is used to predict the suppression of the power spectrum from the baryon fraction of haloes. Conversely, the inverted \texttt{SP(k)} model in \cref{eq:inv_spk} is used to predict the baryon fraction--halo mass relation from a given suppression of $P(k)$.}
 \label{fig:cartoon}
\end{figure}

\section{The SP(k) model}\label{sec:spk}
\texttt{SP(k)} is a parametric model that describes the effects of baryon physics on the non-linear matter power spectrum (namely its suppression as a function of comoving wavenumber, $k$) based on the median baryon fraction of haloes as function of halo mass. The model uses an \textit{optimal mass}, $\hat{M}_{k}$, defined as the halo mass that maximises the strength of the correlation between the suppression of the total matter power spectrum and the total baryon fraction at a given wavenumber $k$. \cite{salcido_2023} provide the following parametric fit for the \textit{optimal mass} based on the ANTILLES simulations:
\begin{equation}\label{eq:optimal_mass_fit}
    \log_{10}\qty(\hat{M}_{k,\mathrm{SO}}(k,z)) = \alpha(z)  - \qty[\alpha(z) - \beta(z)] k^{\gamma(z)} \ \ ,
\end{equation}
where the halo mass $\hat{M}_{k}$ could be specified using either of two different spherical overdensity (SO) masses, namely $M_{200c}$ or $M_{500c}$, which are the masses enclosed within the radius whose mean density is 200 or 500 times the critical density of the Universe, respectively. The functions $\alpha(z)$, $\beta(z)$ and $\gamma(z)$ are modelled with a polynomial fit for their redshift dependence as:
\begin{align}\label{eq:z_dep}
\begin{split}
    X(z) = \sum_{i=0}^2{X_{i}(1+z)^{i}} \ \ ,
\end{split}
\end{align}
where $X=\{\alpha, \beta, \gamma\}$. The best fit coefficients, $X_{i}$, are given in Table 3 in \cite{salcido_2023}. 

The fractional impact of baryons on the total matter power spectrum is then modelled with the function:
\begin{equation}\label{eq:sup_fit}
    P_\mathrm{hydro}(k)/P_\mathrm{DM}(k) = \lambda(k,z) - \qty[\lambda(k,z) - \mu(k,z)] \exp[{-\nu(k,z)\tilde{f}_b}] \\ ,
\end{equation}
where $\tilde{f}_b$ is the baryon fraction at the optimal halo mass normalised by the universal baryon fraction, i.e., 
\begin{equation}\label{eq:opt_fb}
    {\tilde{f}_b = f_b(\hat{M}_{k,\mathrm{SO}}(k,z))/\qty(\Omega_b/\Omega_m)} \ \ ,
\end{equation}
and $\lambda(k,z)$, $\mu(k,z)$, and $\nu(k,z)$ are given by the functional forms:
\begin{align}
    \lambda(k, z) &= 1 + \lambda_a(z) \exp(\lambda_b(z)  \log_{10}(k)) \ \ , \label{eq:func_forms1}\\
    \mu(k, z) &= \mu_a(z) + \frac{1 - \mu_a(z)}{1 + \exp({\mu_b(z) \log_{10}(k) + \mu_c(z)})} \ \ , \label{eq:func_forms2}\\
    \nu(k, z) &= \nu_a(z) \exp(-{\frac{\qty(\log_{10}(k) - \nu_b(z))^2}{2\nu_c(z)^2}}) \ \ . \label{eq:func_forms3}
\end{align}

The evolution of each parameter is modelled as a polynomial function in redshift using \cref{eq:z_dep}, with ${X=\left\{\lambda_a, \lambda_b, \mu_a, \mu_b, \mu_c, \nu_a, \nu_b, \nu_b\right\}}$ accordingly. The best-fit coefficients are given in Table 4 in \cite{salcido_2023}.

\begin{figure}
\centering 
\includegraphics[width=0.48\textwidth]{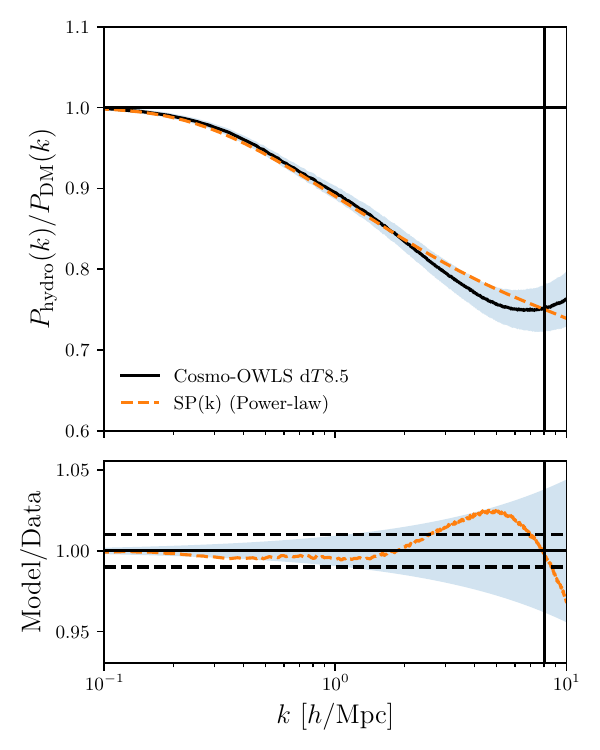}
 \vspace{-1.5em}
 \caption{Fractional impact of baryons on the total matter power spectrum for the cosmo-OWLS d$T8.5$ model \protect\citep{le_brun_towards_2014} at redshift $z=0$. The \texttt{SP(k)} model uncertainties are shown as the light shaded blue region around the true suppression. The best fit model using \texttt{SP(k)} with a simple power-law shape to the $f_b$--$M_\mathrm{halo}$ relation is shown in orange. This simple model can recover the true suppression to within 1\% accuracy for ${k \lesssim 3 h \,\, \mathrm{Mpc}^{-1}}$, but deviates for higher wavenumber. The vertical black line indicates the Nyquist frequency of the simulations $k_{\mathrm{Ny}}$. \textit{Bottom panel:} ratio between the measurements of the suppression in the power spectrum induced by baryons as measured in the simulations to the best-fit simple power-law form for the $f_b$--$M_\mathrm{halo}$ relation in \cref{eq:power-law_fb}. The dashed black lines indicate a 1\% accuracy.}
 \label{fig:PS}
\end{figure}

\begin{figure}
\centering 
\includegraphics[width=0.48\textwidth]{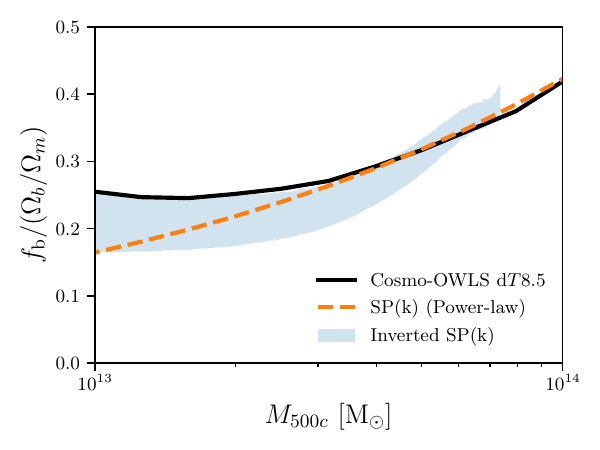}
 \vspace{-1.5em}
 \caption{The $f_b$--$M_\mathrm{halo}$ relation for the cosmo-OWLS d$T8.5$ model \protect\citep{le_brun_towards_2014} at redshift $z=0$. The light shaded blue region shows required baryon fraction to reproduce the power spectrum suppression in \cref{fig:PS} using the inverted \texttt{SP(k)} including the model uncertainties. The true median baryon fraction--halo mas relation lies within the intrinsic model uncertainties. The best fit model for the power spectrum suppression using a simple power-law form to the $f_b$--$M_\mathrm{halo}$ relation is shown in orange. This simple model can recover the true median baryon fraction for $M_{500c} \gtrsim 3 \times {10^{13}\mathrm{M}_\odot}$, but deviates for lower masses.}
 \label{fig:fb}
\end{figure}

\section{Inverted SP(k)}\label{sec:ispk}
A convenient feature of \texttt{SP(k)} is that, because it is based on a set of monotonic analytical equations, the model can be easily ``inverted'' to infer a required baryon fraction--halo mass relation given an input suppression of the power spectrum.  This allows us to quickly check the implied baryon fractions of groups and clusters for a proposed suppression.

Solving for $\tilde{f}_b$ in \cref{eq:sup_fit} yields: 
\begin{equation}\label{eq:inv_spk}
    \tilde{f}_b = \frac{-1}{\nu(k,z)}\ln \qty[\frac{\lambda(k,z) - P_\mathrm{hydro}(k)/P_\mathrm{DM}(k)}{\lambda(k,z) - \mu(k,z)}],
\end{equation}
where the best fitting parameters for $\lambda(k,z)$, $\mu(k,z)$ and $\nu(k,z)$ remain the same as before.

To clarify the mapping between wavenumber $k$ and halo mass $M_{500c}$ (or $M_{200c}$), and how $f_b$ is obtained: for each $k$ and redshift $z$, we use the optimal mass relation in Eq.~(\ref{eq:optimal_mass_fit}) to determine the corresponding $M_{500c}$, i.e., $M_{500c} = \hat{M}_{k,500c}(k,z)$. The solution of Eq.~(\ref{eq:inv_spk}) gives the normalised baryon fraction $\tilde{f}_b$ at this optimal mass. The physical baryon fraction can be recovered as $f_b(M_{500c}) = \tilde{f}_b \times (\Omega_b/\Omega_m)$. This procedure is repeated for each $k$ (and thus $M_{500c}$), yielding the $f_b$--$M_{500c}$ relation. Throughout this work, we always plot the \textit{normalised} baryon fraction, $\tilde{f}_b$, as a function of $M_{500c}$ for consistency with simulation and observational results. For further details, we refer the reader to Section 3.2 of \citet{salcido_2023}.

In \cref{fig:cartoon}, we illustrate how \texttt{SP(k)} and its inverted form in \cref{eq:inv_spk} are used to predict the suppression of the power spectrum based on the baryon fraction of haloes, and vice versa.

We test the accuracy of our inverted \texttt{SP(k)} model to recover the baryon fraction--halo mass relation against the BAHAMAS simulations \citep{mccarthy_bahamas_2017,mccarthy_bahamas_2018}, finding good agreement. We also tested using the cosmo-OWLS simulations \citep{le_brun_towards_2014}, which were not included in either the calibration or validation sets presented in \citet{salcido_2023}. These simulations use a flat $\Lambda$CDM cosmology consistent with the WMAP 7-year results \citep{Komatsu_2011}. The cosmological parameters are \{$\Omega_{m}$, $\Omega_{b}$, $\Omega_{\Lambda}$, $\sigma_{8}$, $n_{s}$, $h$\} = \{0.272, 0.0455, 0.728, 0.81, 0.967, 0.704\}. The simulations consist of a comoving volume with ${400 \,\, \mathrm{cMpc} \, h^{-1}}$ on a side and $2 \times 1024^3$ particles of $m_\mathrm{DM} = 3.75 \times 10^9 \mathrm{M}_\odot \, h^{-1}$ dark matter particle mass, and $m_\mathrm{g} = 7.54 \times 10^8 \mathrm{M}_\odot \, h^{-1}$ initial gas particle mass. A full discussion of the sub-grid implementation, including the prescriptions for star formation, gas heating and cooling, black hole formation, and supernovae and active galactic nuclei (AGN) feedback models can be found in \citet{le_brun_towards_2014} (see also \citealt{schaye_physics_2010}). Due to their feedback implementation, the cosmo-OWLS simulations have lower-than-observed galaxy formation efficiencies for haloes with masses similar to the Milky Way's ($M_{200}\sim10^{12}$ M$_\odot$). Hence, they consistently underpredict the abundance of galaxies with $\log_{10}(M_*/{\rm M}_\odot) < 11$ compared to recent galaxy stellar mass function observations \citep{mccarthy_bahamas_2017}. 

As we will discuss below, we are interested in simulations with strong baryonic suppression as a potential way to reconcile the $S_8$ tension in weak galaxy lensing surveys. Hence, we use the high AGN heating temperature model, cosmo-OWLS d$T8.5$, that predicts a lower gas fraction than inferred from recent X-ray observations (i.e., too much gas ejection). This removal of large quantities of gas results in a suppression of the total matter power spectrum $P(k)$, from large scales of $k \approx 0.1 h \,\, \mathrm{Mpc}^{-1}$ all the way to small scales, $k \gtrsim 10 h \,\, \mathrm{Mpc}^{-1}$ \citep{chisari_modelling_2019,van_daalen_2020}. In \cref{fig:PS} we show the ratio of power spectra for the cosmo-OWLS d$T8.5$ model with respect to its DM-only counterpart at redshift $z=0$. The vertical black line shows the one-dimensional Nyquist frequency of the simulations, $k_{\mathrm{Ny}} = \pi N/L \approx 8 \, {h \,\mathrm{Mpc}^{-1}}$, where $N$ is the cube root of the total number of particles, and $L$ is the length of the box.

It is important to remember that all models have inherent limits and uncertainties associated with them. In particular, \texttt{SP(k)} provides an unbiased estimator of the true baryonic effects for a large ensemble of hydrodynamical simulations, but the error in the model increases with scale $k$, giving rise to a heteroscedastic behaviour \citep{salcido_2023}. We should take these errors into consideration when inverse modelling the required baryon fraction from a given power spectrum suppression. Based on Figs. 8 and 9 in \cite{salcido_2023}, we account for the \textit{maximum} model uncertainties using a simple power-law that goes from a $0.2 \%$ error at $k = 0.1 h \,\, \mathrm{Mpc}^{-1}$ to  $5 \%$ error at $k = 12 h \,\, \mathrm{Mpc}^{-1}$. These uncertainties are shown as the light shaded blue region around the true suppression in \cref{fig:PS}. 

We now use the inverted \texttt{SP(k)} model to compute the required baryon fraction to reproduce the suppression of the matter power spectrum from the cosmo-OWLS d$T8.5$ model. By incorporating the model uncertainties into \cref{eq:inv_spk}, the results are shown as the light shaded blue region in \cref{fig:fb}. The black line shows the true median baryon fraction computed directly from the simulations. We note that, in order to avoid noise in our modelling, we only used \cref{eq:inv_spk} to model scales $k \geq 0.4 \, [h \,\, \mathrm{Mpc}^{-1}]$, as for (very)large scales, the suppression of the matter power spectrum should be close to unity, independent of the baryon fraction. This is because a large fraction of the power on these scales comes from outside $r_{500}$ of haloes \citep{van_daalen_contributions_2015}. Hence, slight variations in the suppression may result in wide variations in the baryon fractions. Because the optimal mass in the \texttt{SP(k)} modelling plateaus for small scales, i.e. a narrow range of the most massive haloes are \textit{mapped} to a significant range of large scales, our choice of `scale cut' does not affect our ability to recover, or predict, the baryon fraction for large haloes \citep[see Fig. 5 in][]{salcido_2023}. A scale cut\footnote{For comparison, a scale cut of $k \geq 0.1 \, [h \,\, \mathrm{Mpc}^{-1}]$ corresponds to an optimal mass of $10 ^{13.89} \mathrm{M}_\odot$} of $k \geq 0.4 \, [h \,\, \mathrm{Mpc}^{-1}]$ corresponds to the largest optimal mass recovered of $10 ^{13.86} \mathrm{M}_\odot$ as shown in \cref{fig:fb}. 

\texttt{SP(k)} is a flexible model that is not restricted to a particular shape of the baryon fraction--halo mass relation, and by inverse modelling this relation using the power spectrum suppression directly, \cref{fig:fb} shows that we are able to recover the true median baryon fraction--halo mas relation within the intrinsic uncertainties. Furthermore, the overall shape of the baryon fraction--halo mass relation form this particular simulation is well recovered.

Additionally, we can test specific functional forms for the baryon fraction--halo mass relation. As an example, we tested a simple power-law functional form for the relation. For the mass range that can be relatively well probed in current X-ray and Sunyaev-Zel'dovich effect observations (roughly $10^{13} \lesssim M_{200\mathrm{c}} \,\, [\mathrm{M}_\odot] \lesssim 10^{15} $), the total baryon fraction of haloes can be roughly approximated by a power-law with constant slope \citep[e.g.][]{Mulroy_2019,Akino_2022}. Hence, we use the following parameterisation,
\begin{equation}\label{eq:power-law_fb}
    f_b / (\Omega_b/\Omega_m) = a \left(\frac{M_{500c}}{10^{13.5}\mathrm{M}_\odot} \right)^{b}, 
\end{equation}
where $a$ is the normalisation of the $f_b-M_{500c}$ relation at a pivot mass of $M_{500c} = {10^{13.5}\mathrm{M}_\odot}$, and $b$ is the power-law slope. 

Using \texttt{SP(k)}, we find the best fitting values of $a$ and $b$ to reproduce the suppression of the matter power spectrum for the cosmo-OWLS d$T8.5$ model directly, rather than fitting the baryon fraction. The best fit model is shown as a dashed orange line in \cref{fig:PS}. The figure shows that a simple power-law functional form can recover the suppression to better than $1\%$ up to $k \approx 3 [h \,\, \mathrm{Mpc}^{-1}]$, and to within $\approx 2\%$ for the entire range shown, up to $k = 10 [h \,\, \mathrm{Mpc}^{-1}]$. This error behaviour with wavenumber for a simple power-law form translates into a  median baryon fraction that agrees with the true median for halo masses $M_{500c} \gtrsim 3 \times {10^{13}\mathrm{M}_\odot}$, but deviates for lower masses, as shown in \cref{fig:fb}. This is expected, as the simulations show a mass-dependent slope for the $f_b$--$M_\mathrm{halo}$ relation.


\section{Results}\label{sec:res}

In this section, we use the inverted form of \texttt{SP(k)}, as developed in \cref{sec:ispk}, to model the baryon fraction--halo mass relation required to reproduce the suppression of the matter power spectrum observed in various cosmic shear analyses. To ensure accurate predictions, we account for the $1\sigma$ uncertainties in the \texttt{SP(k)} model as a function of scale, baryon fraction, and redshift, as described in \citet{salcido_2023}. These uncertainties are incorporated into the tension calculations to avoid overestimating the statistical differences between the inferred and observed baryonic fractions. 

It is important to note that recent studies have demonstrated that the impact of baryonic feedback on the matter power spectrum is not entirely independent of cosmology \citep[e.g.][]{Elbers_2024}. Specifically, the amplitude and shape of the power spectrum suppression can vary slightly depending on the underlying cosmological parameters assumed in the simulations or analyses. As a result, the different cosmologies adopted in the BAHAMAS simulations, the Planck cosmologies assumed in various weak lensing analyses, and the cosmology used by \citet{Akino_2022} could introduce small shifts in the power spectrum. However, these effects are expected to be subdominant compared to the overall trends discussed here and do not qualitatively affect our main conclusions. In particular, \citet{Elbers_2024} found that non-factorizable corrections to the power spectrum due to cosmology dependence are typically below the 1\% level, which is smaller than the statistical and modelling uncertainties considered in this work.

We begin by presenting the results from cosmic shear studies of both the KiDS 1000 and DES Y3 surveys, which marginalise over baryonic effects while holding cosmological parameters fixed to a Planck-like cosmology. We then compare the modelled baryon fractions with the recent baryon budget measurements from \citet{Akino_2022}. In \cref{sec:free_cosmo}, we extend this analysis and comparison to cosmic shear studies that marginalise over both cosmological parameters and baryonic effects. Finally, we examine the findings from the latest DES Y3 weak lensing and kSZ joint analysis in \citet{Bigwood_2024}.

\begin{figure*}
\centering 
\includegraphics[width=0.95\textwidth]{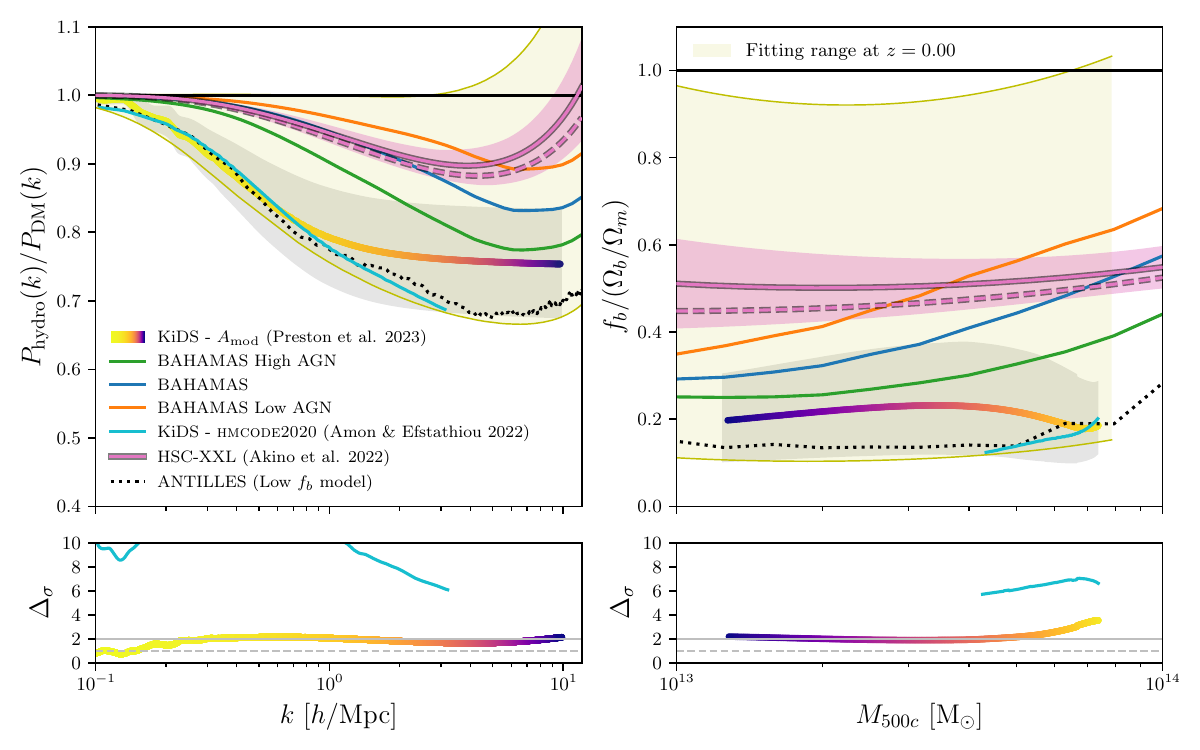}
 \vspace{-1.5em}
 \caption{\textit{Top-Left:} The suppression of the matter power spectrum at redshift $z=0$ required to match the KiDS $\xi_\pm$ shape correlation measurements assuming a Planck $\Lambda$CDM cosmology. The colour line shows the required suppression inferred in \protect\cite{preston_2023} using their phenomenological model $A_\mathrm{mod}$. The line has been colour coded (in both panels) by the \textit{optimal halo mass} that maximises correlation between the baryon fraction and the suppression of the power spectrum (\cref{eq:optimal_mass_fit}). The light grey shaded region encloses the $1\sigma$ confidence interval. The cyan line shows the suppression inferred in \protect\cite{amon_2022} using \textsc{hmcode2020} \protect\citep{mead_hmcode-2020_2020}. The pink line shows the inferred suppression computed using \texttt{SP(k)} based on the median baryon fraction--halo mass relation from the latest HSC-XXL weak lensing and X-ray data from \protect\citet{Akino_2022}, with (solid) and without (dashed) a correction for the contribution of blue galaxies and the diffuse intracluster light. The light pink shaded region enclose the $1\sigma$ confidence interval. For comparison, we show the suppression for three BAHAMAS models (low AGN, fiducial and high AGN \protect\citep{mccarthy_bahamas_2017,mccarthy_bahamas_2018}, as well as one of the most extreme feedback models in the ANTILLES suite \protect\citep{salcido_2023}. The light yellow shaded region shows the entire suppression range spanned in the ANTILLES suite used to train the \texttt{SP(k)} model. \textit{Top-Right:} The $f_b$--$M_\mathrm{halo}$ relation inferred from the suppression of the matter power spectrum on the left panel. The coloured and cyan lines show the baryon fraction inferred using \cref{eq:inv_spk}, required to match the suppression of \protect\cite{preston_2023} using $A_\mathrm{mod}$ and \textsc{hmcode2020} respectively. The pink line shows the the median baryon fraction from \protect\citet{Akino_2022}. We show the baryon fractions measured from the BAHAMAS and ANTILLES simulations for reference. The light yellow shaded region shows the entire range of baryon fractions spanned in the ANTILLES suite used to train the \texttt{SP(k)} model. \textit{Bottom:} Statistical difference between the modelled power spectrum suppression (\textit{Left}) and baryon fractions (\textit{Right}), as compared to the observational results from \protect\citet{Akino_2022}. The solid and dashed grey lines indicate a $2\sigma$ and $1\sigma$ difference respectively.}
 \label{fig:KiDS}
\end{figure*}

\subsection{KiDS 1000 Cosmic shear + Planck \texorpdfstring{$\Lambda$}{Lambda}CDM cosmology.}

In the left panel of \cref{fig:KiDS} we show the power spectrum suppression from \cite{preston_2023} using their phenomenological $A_\mathrm{mod}$ model required to match the KiDS 1000 cosmic shear measurements \citep{Asgari2021} assuming Planck $\Lambda$CDM priors on $S_8$ and $\Omega_m$. The line has been colour coded by the \textit{optimal halo mass} in \cref{eq:optimal_mass_fit}. In cyan, we show the inferred suppression from \cite{amon_2022} using \textsc{hmcode2020} \citep{mead_hmcode-2020_2020}, that fixed the cosmological parameters to the best-fit Planck $\Lambda$CDM values. We note that \cite{amon_2022} did not account for uncertainties on the Planck priors. For clarity, we only show the $1\sigma$ confidence interval for the $A_\mathrm{mod}$ model with a light grey shaded region.

For comparison, we show three feedback variations of the BAHAMAS simulations, Low AGN, Fiducial, and High AGN \citep[with corresponding to $\log_{10} (\Delta T_\mathrm{heat}/K) = \{ 7.6, 7.8, 8.0\}$ respectively,][]{mccarthy_bahamas_2017,mccarthy_bahamas_2018}, where $\Delta T_\mathrm{heat}$ is the BAHAMAS subgrid parameter that controls the temperature increase of gas particles during AGN feedback events. These simulations have been shown to have consistent levels of suppression with the predictions from a joint analysis of KiDS cosmic shear and Sunyaev-Zel'dovich effect data \citep{troster_2022}. We also show for comparison one of the most extreme feedback models in the ANTILLES suite \citep{salcido_2023} in black. The light yellow shaded region shows the entire range of suppression spanned in the ANTILLES suite, which was specifically designed to conservatively bracket current observational constraints (with their associated uncertainties) on the stellar and gas fractions. 

Using \texttt{SP(k)}, we directly model the power spectrum suppression from the observed median baryon fraction of haloes from the latest HSC-XXL weak gravitational lensing data from \citet{Akino_2022}, with (pink solid) and without (pink dashed) a correction for the contribution of blue galaxies and the diffuse intracluster light. The light-shaded region encloses the $1\sigma$ confidence interval. 

As already pointed out by \cite{amon_2022}, for both the $A_\mathrm{mod}$ and the \textsc{hmcode2020} models, a much more aggressive form of feedback beyond what is typically simulated is required to reconcile the KiDS 1000 cosmic shear measurements with a Planck $\Lambda$CDM cosmology. We note that such levels of suppression can be achieved in cosmological hydrodynamical simulations with ``extreme'' feedback prescriptions such as the model shown from the ANTILLES suite (black dotted line). Nevertheless, as we discuss below, such extreme feedback models produce haloes that are highly depleted of their baryons at the present day (see black dotted line in the right panel of \cref{fig:KiDS}), which is at odds with the latest observations of X-ray-selected galaxy groups and clusters \citep[see e.g.][]{Akino_2022}.  

For scales $k>1 \, [h \,\, \mathrm{Mpc}^{-1}]$, the \textsc{hmcode2020} model has similar suppression as the extreme feedback model from the ANTILLES suite. However, the $A_\mathrm{mod}$ model has a suppression shape that plateaus at smaller scales. This feature is not reproduced in simulations as, by construction, the $A_\mathrm{mod}$ phenomenological model only modifies the power spectrum in the non-linear regime.

It should be noted that the suppression inferred using \texttt{SP(k)} for the median baryon fraction of HSC-XXL haloes \citep{Akino_2022} sits somewhere in the middle of the predicted suppression from the Low AGN and the Fiducial BAHAMAS models \citep{mccarthy_bahamas_2017,mccarthy_bahamas_2018}. This is as expected as the BAHAMAS simulations were calibrated specifically to reproduce the gas fraction of observed galaxy groups and clusters. For reasons that will become clear in our discussion below, the \citet{Akino_2022} suppression agrees better with the fiducial BAHAMAS simulation model at large scales $k<1 \, [h \,\, \mathrm{Mpc}^{-1}]$, while the Low AGN model does a better job for scales $k>1 \, [h \,\, \mathrm{Mpc}^{-1}]$. 

In the right panel of \cref{fig:KiDS} we use the inverted form of \texttt{SP(k)} in \cref{eq:inv_spk} to model the required baryon fraction to reproduce the suppression of the matter power spectrum from the KiDS $\xi_\pm$ measurements in the left panel. The $A_\mathrm{mod}$ model has been colour coded by the \textit{optimal halo mass} in the same way as in the left panel. This allows us to directly compare the $k$ scales that are used in the calculation of $f_b$. Note that the mass ranges shown for each model correspond to the scales shown on the left panel. For example, the cyan line is only modelled for halo masses  $M_{500c} \gtrsim 4 \times 10^{13} [\mathrm{M}_\odot]$, as the published power spectrum suppression is only available for scales $k \lesssim 3 \, [h \,\, \mathrm{Mpc}^{-1}]$.   

The figure shows that the baryon fraction of a narrow range of the most massive haloes ($6\times 10^{13} \lesssim M_{500c}\, [\mathrm{M}_\odot] \lesssim 8 \times 10^{13} $) are \textit{mapped} to a significant range of large scales ($0.2 \lesssim k \, [h \,\, \mathrm{Mpc}^{-1}]\lesssim 1$), while a large range of halo masses ($10^{13} \lesssim M_{500c}\, [\mathrm{M}_\odot] \lesssim 6 \times 10^{13} $) are mapped to a comparable range of small, non-linear scales ($1 \lesssim k \, [h \,\, \mathrm{Mpc}^{-1}]\lesssim 10$). It follows from this mapping that the fiducial BAHAMAS simulation model agrees better with an inferred suppression from the median baryon fraction of haloes from HSC-XXL at large scales, as the blue line in the right panel in \cref{fig:KiDS} overlaps with the pink $1\sigma$ region at large halo masses. Similarly, the Low AGN model does a better job for scales $k>1 \, [h \,\, \mathrm{Mpc}^{-1}]$ as the orange line overlaps with the \cite{Akino_2022} data at lower halo masses. 

In order to quantify the statistical difference between the observed and predicted baryon fractions, assuming that the probability density function is approximated by a Gaussian distribution at each halo mass scale, we compute,
\begin{equation}
    \Delta\sigma=\frac{f_{b,\mathrm{obs}} - f_{b, \mathrm{model}}}{\sqrt{\sigma_\mathrm{obs}^2 + \sigma_\mathrm{model}^2}},
\end{equation} 
where we used the associated uncertainties for each quantity. 

\begin{figure*}
\centering 
\includegraphics[width=0.95\textwidth]{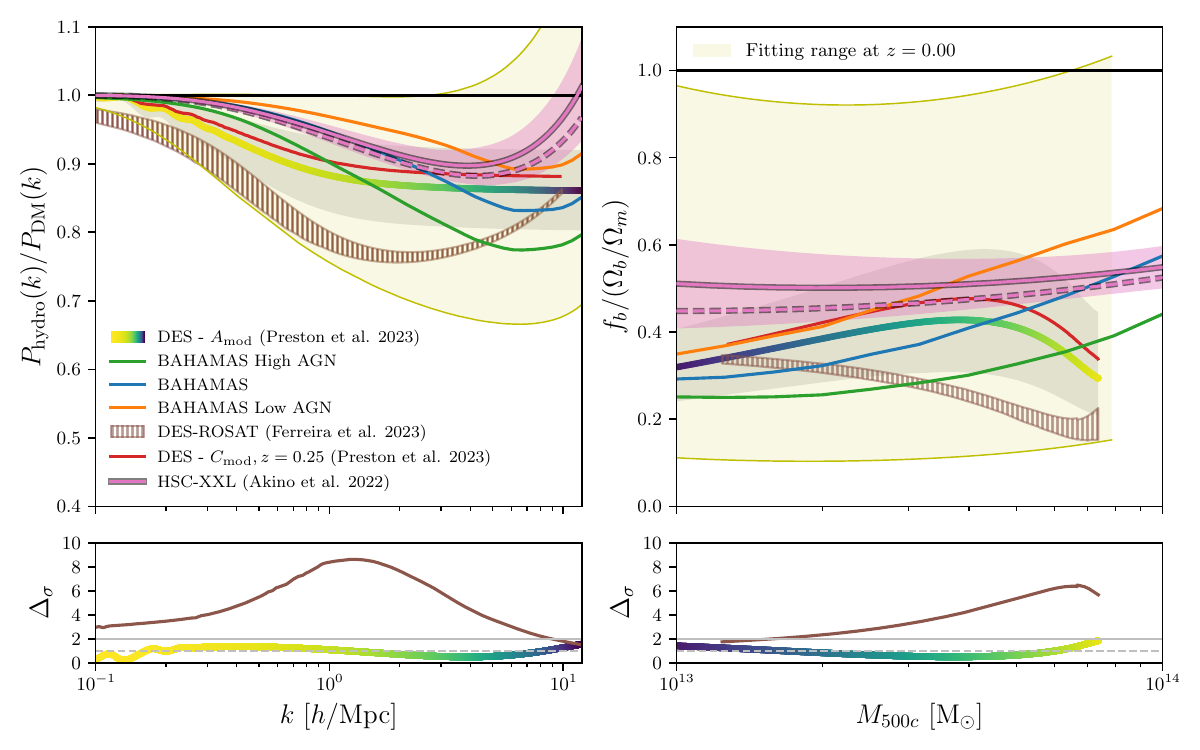}
 \vspace{-1.5em}
 \caption{Same as in \cref{fig:KiDS}, but for the DES Y3 lensing results. \textit{Top-Left:} The colour line shows the required suppression inferred in \protect\cite{preston_2023} using their phenomenological model $A_\mathrm{mod}$. The line has been colour coded (in both panels) by the \textit{optimal halo mass}. The light grey shaded region encloses the $1\sigma$ confidence interval. The red line shows the suppression inferred in \protect\cite{preston_2023} using the six parameter redshift-dependent model $C_\mathrm{mod}$ (only shown at redshift of $z=0.25$ for reference). The brown hatched region shows the recent baryonic effects inferred from the cross-correlation between cosmic shear and the diffuse X-ray background using DES and ROSAT from \protect\cite{Ferreira_2023}. \textit{Top-Right:} The coloured and brown lines show the baryon fraction inferred using \cref{eq:inv_spk}, required to match the suppression of \protect\cite{preston_2023} using $A_\mathrm{mod}$,  $C_\mathrm{mod}$, and \protect\cite{Ferreira_2023}, respectively. \textit{Bottom:} Statistical difference between the modelled power spectrum suppression (\textit{Left}) and baryon fractions (\textit{Right}), as compare to the observational results from \protect\citet{Akino_2022}. The solid and dashed grey lines indicate a $2\sigma$ and $1\sigma$ difference respectively.}
 \label{fig:DESY3}
\end{figure*}

Similarly, for the power spectrum suppression ${S(k) = P_\mathrm{hydro}(k)/P_\mathrm{DM}(k)}$, we use, 
\begin{equation}
    \Delta\sigma=\frac{S(k)_{\mathrm{obs}} - S(k)_{\mathrm{model}}}{\sqrt{\sigma_\mathrm{obs}^2 + \sigma_\mathrm{model}^2}}.
\end{equation} 
In both cases, the subscript ‘\textit{obs}’ refers to the observed baryon fraction of haloes from the latest HSC-XXL weak gravitational lensing and X-ray data from \citet{Akino_2022}, or its corresponding modelled suppression using \texttt{SP(k)}. We show this measurement of statistical “tension” for $f_b$ and the power spectrum suppression in the bottom panels of \cref{fig:KiDS,fig:DESY3,fig:DESY3_kSZ}. These tension values account for the internal uncertainties of the SP(k) model as described in \citet{salcido_2023}.

The bottom right panel shows that the $A_\mathrm{mod}$ model predicts baryon fractions that are more than $2 \sigma$’s away from the mean observed baryon fraction in \cite{Akino_2022} in many adjacent bins. As we do not have access to the full covariance matrix of these studies, we cannot calculate a global measure of tension, but we expect that summing over the bins (while accounting for correlations between them) would increase the statistical tension between the datasets. Furthermore, for masses of $M_{500c} \gtrsim 5 \times 10^{13} \, [\mathrm{M}_\odot]$, the $A_\mathrm{mod}$ model predicts baryon fractions that mildly decrease with increasing halo mass. This behaviour is not reproduced in the observations or in simulations, and is a consequence of $A_\mathrm{mod}$ only modifying the power spectrum in the non-linear regime.  At scales of $\approx1 \, [h \,\, \mathrm{Mpc}^{-1}]$, which is the range that current cosmic shear data is most sensitive to, the level of statistical tension with respect to the observed baryon fraction of up to ${\approx} 3.5 \sigma$ for halo masses between $7 \times 10^{13} \lesssim M_{500c} \, [\mathrm{M}_\odot] \lesssim 8  \times 10^{13}$. 

As the \textsc{hmcode2020} model was, to an extent, calibrated on hydrodynamical simulations, the power spectrum suppression has a similar shape and amplitude to that of the ANTILLLES “extreme” feedback model (left panel). Consequently, for such extreme suppression, both models show an extremely low baryon fraction, where haloes are almost entirely depleted of their baryons (right panel). This severe baryon deficiency is in strong tension with the mean observed baryon fraction of \cite{Akino_2022}. While the \textsc{hmcode2020} model has a similar level of baryonic suppression as the $A_\mathrm{mod}$ model for large scales, the increased tension is a result of the smaller uncertainties for the \textsc{hmcode2020} model. 

The bottom left panel of \cref{fig:KiDS} shows that the $A_\mathrm{mod}$ model is within ${\approx} 2\sigma$ from the the suppression inferred using \texttt{SP(k)} for the median baryon fraction of haloes from HSC-XXL \citep{Akino_2022}. The slightly better statistical agreement in the suppression compared to the baryon fraction, especially at high halo masses,  is due to the non-linear mapping between the baryon fraction and the power spectrum suppression.

Based on \cref{fig:KiDS}, we generally conclude that, while invoking an aggressive form of baryonic feedback could in principle reconcile the primary CMB(+BAO+CMB lensing) measurements with low-redshift LSS measurements, this introduces another tension with the observed median baryon fraction of haloes.  In Section \ref{sec:Conclusions} we discuss possible ways to avoid this new tension. 

\begin{figure*}
\centering 
\includegraphics[width=0.95\textwidth]{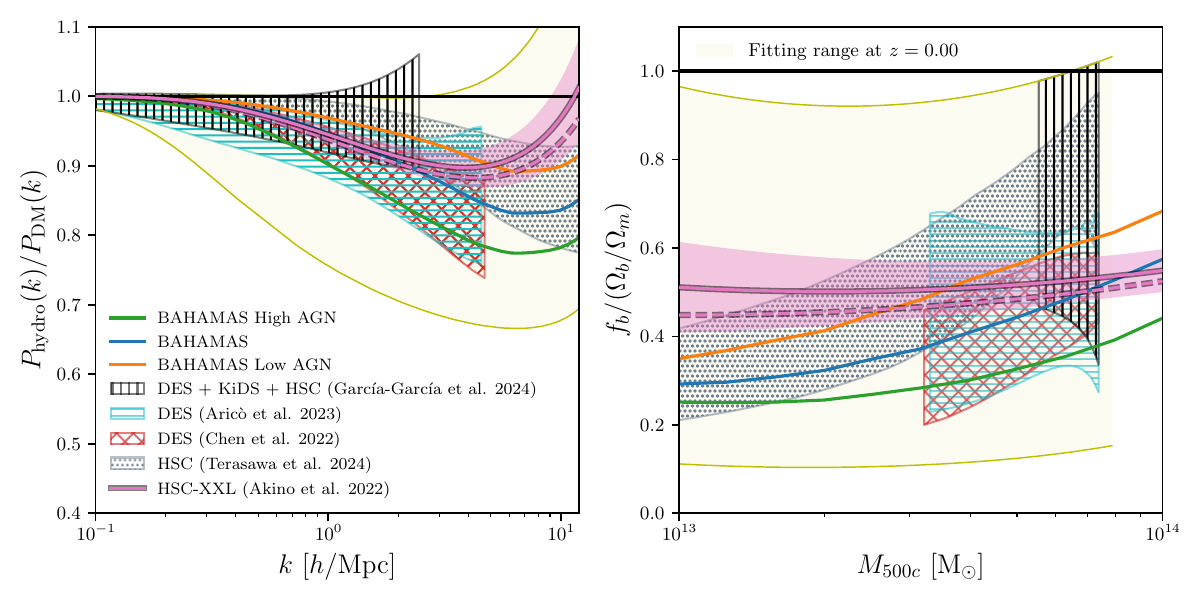}
 \vspace{-1.5em}
 \caption{Same as in \cref{fig:KiDS,fig:DESY3,fig:DESY3_kSZ} but for weak lensing analyses that leave the cosmological parameters free. We show the suppression inferred in \protect\cite{arico_2023}, \protect\cite{chen_2023}, and \protect\cite{Garcia_2024} using the \texttt{BACCOemu} \protect\cite{arico_2021} emulator, and \protect\cite{Terasawa_2024} using the \protect\cite{mead_hmcode-2020_2020} halo model. The cyan, red, black and grey hatched regions show the 68 per cent credible region for the suppression inferred in \protect\citet[][DES Y3]{arico_2023}, \protect\citet[][DES Y3]{chen_2023}, \protect\citet[][DES Y3 + KiDS 1000 + HSC-DR1]{Garcia_2024} and \protect\citet[][HSC-Y3]{Terasawa_2024} respectively. On the right panel we show the corresponding baryon fraction inferred using the inverted for of \texttt{SP(k)} in \cref{eq:inv_spk}. All the models shown here have baryon fractions that are statistically consistent with the observational results from \protect\citet{Akino_2022}.}
 \label{fig:other_studies}
\end{figure*}

\subsection{DES Y3 Cosmic shear + Planck \texorpdfstring{$\Lambda$}{Lambda}CDM cosmology.}
\Cref{fig:DESY3} follows the same format as \cref{fig:KiDS}, but for the the DES Y3 cosmic shear measurements \citep{Amon2022,Secco2022}. We show the power spectrum suppression from \cite{preston_2023} required to match the DES Y3 cosmic shear measurements assuming Planck $\Lambda$CDM priors on $S_8$ and $\Omega_m$ and using their $A_\mathrm{mod}$ model. The line has been color coded by the \textit{optimal halo mass} in \cref{eq:optimal_mass_fit}. The red line shows the results using their six parameter redshift-dependent model $C_\mathrm{mod}$, only shown at the published redshift of $z=0.25$ for reference. For clarity, we only show the $1\sigma$ confidence interval for the $A_\mathrm{mod}$ model with a light grey shaded region.

The brown hatched region shows the recent baryonic effects inferred in \cite{Ferreira_2023}, using the cross-correlation between the DES cosmic shear and the diffuse X-ray background from ROSAT \citep{ROSAT_1999}. \cite{Ferreira_2023} used a halo model similar to that of \cite{mead_hydrodynamical_2020} to model baryonic effects, with ‘bloating’ parameters that modulates the concentration of dark matter haloes, but keeping all cosmological parameters fixed to the best-fit Planck cosmology.

For comparison, we also show three feedback variations of the BAHAMAS simulations and the power spectrum suppression modelled using \texttt{SP(k)} for the observed baryon fraction from \citet{Akino_2022}.

The figure shows that, using both the $A_\mathrm{mod}$ and $C_\mathrm{mod}$ models in \cite{preston_2023}, the power spectrum suppression due to baryons required to match DES Y3 cosmic shear measurements with a Planck cosmology is not as strong as for the KiDS 1000 measurements. However, the results from \cite{Ferreira_2023} using the cross-correlation between DES cosmic shear and the diffuse X-ray background from ROSAT, require a stronger suppression, similar to that required to reconcile the KiDS 1000 measurements.

In the right panel of \cref{fig:DESY3} we use the inverted form of \texttt{SP(k)} in \cref{eq:inv_spk} to predict the required baryon fraction to reproduce the suppression of the matter power spectrum from the DES Y3 $\xi_\pm$ measurements in the left panel. Similar to the behaviour shown in \cref{fig:KiDS}, for masses of $M_{500c} \gtrsim 5 \times 10^{13} \, [\mathrm{M}_\odot]$, both the $A_\mathrm{mod}$ and the redshift-dependent $C_\mathrm{mod}$ models predict baryon fractions that decrease with increasing halo mass. On the other hand, the modelled baryon fraction from the \cite{Ferreira_2023} suppression, shows a baryon fraction that decreases with halo mass. While some simulations within the ANTILLES suite show a similar behaviour, this is at odds with the increasing ‘power-law-like’ behaviour of the observed baryon fraction--halo mass relation \citep[see e.g.][]{Akino_2022}. This discrepancy may be a consequence of the parameter choices and their associated priors in \cite{Ferreira_2023}. 

We have verified that all the power spectrum suppressions and associated baryon fractions modelled in this study, including those from $A_\mathrm{mod}$ and DES-ROSAT (\citealt{Ferreira_2023}), lie within the parameter space spanned by the ANTILLES suite, as indicated by the light yellow shaded region in \cref{fig:DESY3}. The \texttt{SP(k)} fits to these suppression curves are well-behaved and the best-fit parameters remain within the training domain of the model, ensuring that the mapping is robust and not an extrapolation.

To build further intuition for the mapping between power spectrum suppression and baryon fraction, we note that the shape of the suppression curve from \cite{Ferreira_2023} (left panel) is distinct compared to the simulation-based models. At the largest scales, the \cite{Ferreira_2023} suppression exhibits a much stronger suppression than any of the simulation examples, including the BAHAMAS high AGN model (green). Since large-scale power is dominated by the most massive haloes, this translates to a very low baryon fraction at high halo masses in the right panel, as expected from the established correlation between baryon fraction and $P(k)$ suppression \citep{salcido_2023,van_daalen_contributions_2015}. As we move to smaller scales, the \cite{Ferreira_2023} suppression decreases rapidly, crossing over the BAHAMAS high AGN (green) and fiducial (orange) models, and eventually reaching a similar level of suppression as the DES $A_\mathrm{mod}$ results (Preston et al. 2023) at $k \sim 9\,h\,\mathrm{Mpc}^{-1}$. This behaviour maps to a baryon fraction that increases from right to left (i.e., from high to low halo mass), again crossing over the BAHAMAS models and matching the $A_\mathrm{mod}$-inferred baryon fraction at $M_{500c} \lesssim 2\times10^{13}\,\mathrm{M}_\odot$. This is consistent with the fact that lower-mass haloes contribute more strongly to the power spectrum at smaller scales \citep{salcido_2023,van_daalen_contributions_2015}. Thus, the unusual trend in the inferred baryon fraction for the \cite{Ferreira_2023} suppression is a direct consequence of the scale-dependent shape of the input suppression curve, rather than an artefact of the \texttt{SP(k)} model or its training domain.

Interestingly, applying \texttt{SP(k)} to the DES-ROSAT suppression from \citet{Ferreira_2023} results in a decreasing baryon fraction with increasing halo mass, in contrast to the halo model in Ferreira et al., which explicitly enforces a bound baryon fraction that increases with halo mass. This indicates that the mapping between power spectrum suppression and baryon fraction may differ substantially between the Ferreira et al. approach and the ANTILLES simulations. One possible explanation is differences in gas profiles beyond $R_{500}$, as demonstrated by \citet{debackere_2021}, which can introduce significant scatter in this mapping. A detailed comparison of the underlying halo profiles would help clarify this discrepancy, but we leave such an investigation to future work.

The bottom right panel shows that an “$A_\mathrm{mod}$-like” suppression required to reconcile DES Y3 cosmic shear measurements with Planck CMB measurements, would only be in mild $\approx 1.5 - 2.0 \sigma$ tension, particularly at halo masses between $7 \times 10^{13} \lesssim M_{500c} \, [\mathrm{M}_\odot] \lesssim 8  \times 10^{13}$. On the other hand, because of the decreasing baryon fraction with halo mass, in contrast with the increasing behaviour in observations, the \cite{Ferreira_2023} model shows an increasing statistical tension with halo mass, reaching up to $\approx 6.5 \sigma$ at halo masses of $6 \times 10^{13} \lesssim M_{500c} \, [\mathrm{M}_\odot] \lesssim 8 \times 10^{13}$. 

Based on \cref{fig:DESY3}, we can conclude that, while not as extreme as in the case of KiDS 1000, invoking a more aggressive form of baryonic feedback may reconcile the primary CMB(+BAO+CMB lensing) measurements with low-redshift LSS measurements from DES, but this will introduce a mild tension with the observed median baryon fractions of X-ray-selected groups and clusters. This is especially true for large, quasi-linear scales ($k \leq 0.3 h \,\, \mathrm{Mpc}^{-1}$), where the $A_\mathrm{mod}$  model requires a large suppression that translates into a significantly lower baryon fraction compared to observations at relatively large halo masses of $7 \times 10^{13} \lesssim M_{500c} \, [\mathrm{M}_\odot] \lesssim 8  \times 10^{13}$.  This is also the regime where current cosmic shear data is most sensitive.

The cross-correlation between DES cosmic shear and the diffuse X-ray background from ROSAT, as presented in \cite{Ferreira_2023}, requires a stronger suppression, similar to that required to reconcile the KiDS 1000 measurements. Hence, this will introduce another tension with the observed median baryon fraction of haloes.


\subsection{Cosmic shear + Free cosmology}\label{sec:free_cosmo}

In \cref{fig:other_studies} we compare the baryonic feedback constraints from cosmic shear measurements that marginalise over both cosmological parameters and baryonic effects using the using the \texttt{BACCOemu} emulator \citep{Angulo_2021_BACCO,arico_2021,arico_2021_Simultaneous}, also based on the baryonification model \citep{schneider_2015}. While both \citet[][cyan]{arico_2023} and \citet[][red]{chen_2023} used the small-scale DES Y3 shear measurements to constrain baryonic effects, \cite{chen_2023} only vary one \texttt{BACCOemu} baryonic parameter, namely, the parameter that controls the characteristic halo mass in which half of the cosmic gas fraction is expelled from the halo ($M_c$), whereas \cite{arico_2023} set free all the \texttt{BACCOemu} baryonic feedback parameters, i.e. the parameters controlling the shape of the density profile of the hot gas, the galaxy-halo mass ratio, the AGN feedback range, and the gas fraction--halo mass slope. We show in black, the recent analysis of \cite{Garcia_2024} that combines the DES Y3, KiDS 1000 and the Hyper Suprime-Cam \citep[][HSC-DR1]{Aihara_2018_data_release} weak lensing samples under a joint harmonic-space. The authors also use the \texttt{BACCOemu} emulator to marginalise over baryonic effects. Finally, the grey region shows the analysis of \cite{Terasawa_2024} that used the \cite{mead_hmcode-2020_2020} halo model to explore the baryonic effect signature in the HSC Y3 cosmic shear data \citep{L1_2022}. The hatched areas enclose the 68 per cent credible region for each study. As before, we also show three feedback variations of the BAHAMAS simulations and the power spectrum suppression modelled using \texttt{SP(k)} for the observed baryon fraction from \citet{Akino_2022} for comparison.

In the right panel of \cref{fig:other_studies} we use the inverted form of \texttt{SP(k)} in \cref{eq:inv_spk} to model the required baryon fraction to reproduce the suppression of the matter power spectrum from each study. We note that upper limit for the $P(k)$ suppression inferred in \citet{Garcia_2024} falls outside the fitting range of \texttt{SP(k)} for ${k \gtrsim 1 h \,\, \mathrm{Mpc}^{-1}}$. We do not extrapolate our model outside its fitting range, hence, the inferred baryon fraction for \citet{Garcia_2024} using the inverted form of \texttt{SP(k)} hit the upper fitting limit for most of the modelled range.

The figure shows that the inferred baryon fractions from the \citet{arico_2023}, \citet{chen_2023}, \citet{Garcia_2024} and \citet{Terasawa_2024} studies, which marginalise over cosmological parameters and baryonic effects without imposing a Planck-like cosmology, are statistically compatible with the observed baryon budget from the HSC-XXL weak gravitational lensing data from \citet{Akino_2022}. 

Comparing these results with the previous sections, it is clear that marginalising over baryonic effects in weak lensing studies, while keeping the cosmological parameters consistent with the latest Planck measurements of the primary CMB(+BAO+CMB lensing), leads to a stronger inferred $P(k)$ suppression to compensate for the higher values of $S_8$ and $\Omega_m$ preferred by Planck \citep[see also][]{mccarthy_bahamas_2018,Garcia_2024,Terasawa_2024}.

\begin{figure*}
\centering 
\includegraphics[width=0.95\textwidth]{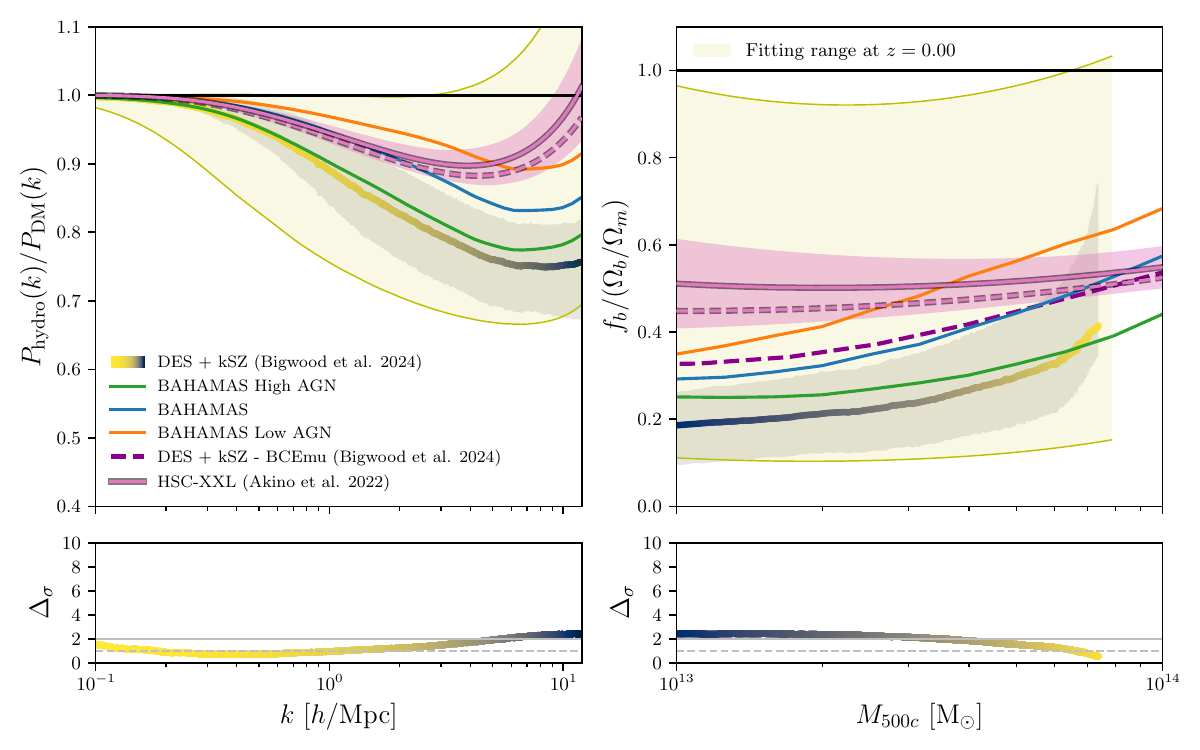}
 \vspace{-1.5em}
 \caption{Same as in \cref{fig:DESY3}, for the joint analysis of the DES Y3 Cosmic shear with the ACT DR5 kinetic Sunyaev Zel'dovich effect. \textit{Top-Left:} The colour line shows the required suppression inferred in \protect\cite{Bigwood_2024} using the \texttt{BCEmu} emulators based on the baryonification model \protect\citep{schneider_2015,Giri_2021}. The line has been colour coded (in both panels) by the \textit{optimal halo mass}. The light grey shaded region encloses the $1\sigma$ confidence interval. \textit{Top-Right:} The coloured line show the baryon fraction inferred using \cref{eq:inv_spk}, required to match the suppression of \protect\cite{Bigwood_2024}. The dashed purple line shows the inferred baryon suppression directly inferred from \texttt{BCEmu}, as presented in \protect\cite{Bigwood_2024}. \textit{Bottom:} Statistical difference between the modelled power spectrum suppression (\textit{Left}) and baryon fractions (\textit{Right}), as compare to the observational results from \protect\citet{Akino_2022}. The solid and dashed grey lines indicate a $2\sigma$ and $1\sigma$ difference respectively.}
 \label{fig:DESY3_kSZ}
\end{figure*}

\subsubsection{DES Y3 Cosmic shear + kinetic Sunyaev Zel'dovich effect.}

Recently, \cite{Bigwood_2024} presented a joint analysis of the DES Y3 Cosmic shear weak lensing data with the Atacama Cosmology Telescope DR5 \citep[][]{ACT_2020} measurements of the kinetic Sunyaev Zel'dovich effect in \cite{Schaan_2021}. Their analysis jointly constrained cosmological and astrophysical baryonic feedback parameters using the \texttt{BCEmu} emulators based on the \textit{‘baryonification’} model \protect\citep{schneider_2015,Giri_2021}. We show the power spectrum suppression from \cite{Bigwood_2024} color coded by the \textit{optimal halo mass} in \cref{eq:optimal_mass_fit}. The $1\sigma$ confidence interval is shown with a light grey shaded region. We show three feedback variations of the BAHAMAS simulations and the power spectrum suppression modelled using \texttt{SP(k)} for the observed baryon fraction from \citet{Akino_2022} for comparison.

The figure shows that the latest weak lensing and kinetic Sunyaev-Zel'dovich analysis favours a slightly higher baryonic feedback suppression, similar to that of the high AGN feedback model in the BAHAMAS simulation suite. The inclusion of kSZ effect data does not appear to be the primary factor behind the differences observed between the results of \cite{Bigwood_2024}, \cite{arico_2023}, \cite{chen_2023}, and \cite{Garcia_2024}. Notably, \cite{Bigwood_2024} reports a somewhat stronger suppression in the matter power spectrum from WL alone compared to \cite{arico_2023,chen_2023}, or \cite{Garcia_2024}, suggesting that the differences are more influenced by the WL data and associated modelling choices (e.g., adopted prior ranges on the baryonic modelling) than by the incorporation of the kSZ effect.  Nevertheless, the differences are not highly statistically significant, particularly on large scales ($k<1 \, [h \,\, \mathrm{Mpc}^{-1}]$) that dominate the lensing signal to noise.

In the top right panel of \cref{fig:DESY3_kSZ} we use the inverted form of \texttt{SP(k)} in \cref{eq:inv_spk} to model the required baryon fraction to reproduce the suppression of the matter power spectrum from the DES Y3 weak lensing + kSZ joint analysis in the left panel. Since the level of power spectrum suppression in the WL + kSZ analysis is similar to that of the high AGN BAHAMAS model, as expected, both have similar baryon fractions. 

The bottom right panel shows the level of statistical tension between the baryon fraction inferred from the DES Y3 weak lensing + kSZ joint analysis and the observed baryon fraction--halo mass relation. The figure show that the results from \cite{Bigwood_2024} would be in mild $\approx 2 - 2.5 \sigma$ tension for halo masses $M_{500c} \lesssim 5 \times 10^{13} \, [\mathrm{M}_\odot]$, but with a lower degree of tension at higher masses.

In the top right panel of \cref{fig:DESY3_kSZ}, we also show with a dashed purple line the baryon fraction--halo mass relation directly modelled from \texttt{BCEmu}. We note the significant difference in the inferred baryon fraction for the same power spectrum suppression using either \texttt{SP(k)} or \texttt{BCEmu}. This discrepancy was already highlighted in \citet[][see their appendix B4]{Bigwood_2024}. Recent studies have confirmed a strong correlation between the predicted impact of baryons on the present-day $P(k)$ from different simulations and the baryon fraction of groups and clusters of such simulations \cite{van_daalen_2020,salcido_2023}. The \texttt{SP(k)} model directly exploits this strong correlation, while the \texttt{BCEmu} model does not enforce a relationship between the baryon fraction and matter power suppression. Therefore, the \texttt{BCEmu} model can predict a more extreme matter power spectrum suppression for a given mean baryon fraction, beyond what has been found in cosmological hydrodynamical simulations \citep[see also appendix B4 in][]{Bigwood_2024}. It is important to note, however, that \texttt{BCEmu} does include cosmology dependence through the baryon fraction parameter, $f_b = \Omega_b / \Omega_m$, which is treated as a free parameter in the emulator \citep{Giri_2021}. Thus, the predicted suppression from \texttt{BCEmu} is not entirely independent of cosmology. 

While the baryonification formalism can provide a high level of flexibility, it is still unclear if some of the parameter space may produce unrealistic results due to the lack of self-consistency of the method (e.g., the gas profiles are specified without regard for the evolutionary history of the halo and the energetics required to modify the profiles in the specified way). On the other hand, cosmological hydrodynamical simulations may not explore the full range of physical possibilities. For instance, while \cite{debackere_2021} have shown that the behaviour of the profiles between $r_{500}$ and $r_{200}$ can affect the matter power spectrum if the profiles are allowed to vary significantly over this range, the effect of baryons could be underestimated if gas is ejected much further away than normally found in hydrodynamical simulations \citep{Garcia_2024}. Hence, further work is essential to test the robustness of the correlation between the power spectrum suppression and the baryon fraction, and to provide physical priors to flexible methods such as the halo model and the baryonification formalism. 

\section{Summary and Conclusions}\label{sec:Conclusions}

While the origin of the ‘$S_8$ tension’ between high-redshift CMB and low-redshift LSS measurements remains unclear, recent studies suggest that baryonic effects alone may be insufficient to address this discrepancy \citep{mccarthy_bahamas_2018,McCarthy_2023}. Others propose that incorporating mechanisms that strongly suppress the non-linear power spectrum could help reconcile these observations \citep{Amon2022,preston_2023}. In this study, we presented a novel method to model the required baryon fraction-halo mass relation from any power spectrum suppression inferred from weak lensing studies, allowing us to test such studies against the measured low-redshift baryon budget estimates in galaxy groups and clusters.

Our specific findings can be summarised as follows:

\begin{itemize} 
    \item We introduced an inverted form of the analytical \texttt{SP(k)} model \citep{salcido_2023}, which enables the computation of the baryon fraction-halo mass relation required to produce a given suppression of the power spectrum.

    \item By employing this inverted \texttt{SP(k)} model, we calculated the baryon fractions necessary to replicate the suppression of the matter power spectrum observed in various cosmic shear and cosmic shear cross-correlation analyses. 

    \item Studies that marginalise over baryonic effects while either holding cosmological parameters fixed to a Planck-like cosmology or jointly fitting to Planck, such as those by \cite{Amon2022}, \cite{Ferreira_2023}, and \cite{preston_2023}, predict a strong suppression of the power spectrum, $P(k)$, to compensate for the higher values of $S_8$ and $\Omega_m$ favoured by Planck.  The inferred baryon fractions from these studies are significantly lower than those measured by the latest HSC-XXL X-ray and weak gravitational lensing data from \cite{Akino_2022}.  

    \item The suppression inferred using the “$A_\mathrm{mod}$” model \citep{Amon2022} required to reconcile KiDS 1000 cosmic shear measurements with Planck CMB measurements would introduce a ${\gtrsim} 2 \sigma$ tension with the observed baryon fraction of galaxy groups and clusters, and up to a ${\approx} 3.5 \sigma$ tension for halo masses in the range $7 \times 10^{13} \lesssim M_{500c} , [\mathrm{M}_\odot] \lesssim 8 \times 10^{13}$.

    \item For DES Y3 cosmic shear measurements, the “$A_\mathrm{mod}$-like” suppression required to reconcile them with Planck CMB measurements would result in only a mild $\approx 1.5 - 2.0 \sigma$ tension with the observed baryon fraction of groups and clusters.

    \item The results from \cite{Ferreira_2023} using the cross-correlation between DES cosmic shear and the diffuse X-ray background from ROSAT using a fixed Planck cosmology introduce a statistical tension with the observed baryon fraction reaching up to $\approx 6.5 \sigma$ at halo masses of $6 \times 10^{13} \lesssim M_{500c} \, [\mathrm{M}_\odot] \lesssim 8 \times 10^{13}$.

    \item In contrast, studies that marginalise over both cosmological parameters and baryonic effects, such as \cite{arico_2021}, \cite{chen_2023}, \cite{Garcia_2024}, and \cite{Terasawa_2024}, still exhibit the $S_8$ tension but yield baryon fractions that are in good statistical agreement with the observations reported by \citet{Akino_2022}. 

    \item The WL+kSZ prediction of a more extreme suppression of the matter power spectrum in \cite{Bigwood_2024} suggests a mild $\approx 2 - 2.5 \sigma$ tension with the observed baryon fraction in haloes with mass $M_{500c} \lesssim 5 \times 10^{13} \, [\mathrm{M}_\odot]$, with a smaller tension at higher masses.

    \item Both the 'bloating' of dark matter halo concentrations via the modified halo model \citep{Mead_16}, as applied in \cite{Ferreira_2023}, and the use of the baryonification model \citep{schneider_2015}, as employed in \cite{Bigwood_2024}, can produce baryon fraction--halo mass relations that differ notably from those predicted by full hydrodynamical simulations. These discrepancies may lead to tensions of several $\sigma$ with the observed baryon fractions. Consequently, further work is necessary to establish more physically motivated priors for flexible approaches like the halo model and baryonification formalism.

\end{itemize}

This study underscores the importance of carefully incorporating baryonic effects in cosmological pipelines for weak lensing studies, as well as ensuring consistency with other physically correlated observables, including the baryon fractions of galaxy groups and clusters \citep[see also ][]{Bigwood_2024}. 

Taking the observed baryon fractions of X-ray-selected groups and clusters at face value and under the assumption that the \texttt{SP(k)} model provides a realistic mapping between the matter power spectrum and baryon fractions, our results suggest that a mechanism beyond baryonic physics alone is required to modify or slow the growth of structure in the universe to resolve the $S_8$ tension.  Our findings therefore suggest that another mechanism (e.g., new dark sector physics) may be at work and/or that there are unaccounted for systematic errors in the lensing or CMB measurements.

Alternatively, if feedback is the principle driver of the $S_8$ tension, the implication would be that current estimates of the baryon fraction--halo mass relation are strongly biased high, by up to a factor of several.  This seems unlikely but would be a stunning development for our understanding intracluster medium if true.  Note that previous work based on forward modelling of simulations has shown that observational methods such as those employed in \citet{Akino_2022} are unbiased on average at the few percent level for weak lensing-based halo mass estimates (e.g., \citealt{becker_2011,bahe_2012}) and X-ray-based gas mass estimates (e.g, \citealt{nagai_2007,le_brun_towards_2014}).  Thus, there are unlikely to be large systematic errors in the observational baryon fraction estimates.  However, correcting for the effects of X-ray-selection is more challenging, particularly at the group scale (e.g., \citealt{pearson_2017,andreon_2024,marini_2024}).  If large numbers of gas-deficient groups are present in nature and unaccounted for the X-ray selection modelling, it may be possible to find consistency with the suppression in $P(k)$ required to reconcile cosmic shear measurements with the primary CMB.  Note, however, that such an explanation is unlikely to be successful for reconciling the similar offsets reported between the CMB and the tSZ power spectrum and its cross-correlation with the cosmic shear \citep{McCarthy_2023}.  That is because these measures of clustering are sensitive to the most massive haloes where the luminosities and baryon fractions are considerably higher than in groups and the selection effects are much better understood.

Recent results from the kinematic Sunyaev-Zel’dovich (kSZ) effect provide compelling evidence of a more extended gas distribution around dark matter haloes \citep{Hadzhiyska_2024}, strongly disfavoring hydrodynamical simulations with weak feedback models. Conversely, the recent detection of the patchy screening effect, which probes the distribution of electrons around galaxies, supports the idea of extended gas distributions but aligns better with simulations that predict less heating and redistribution of the intracluster medium than those suggested by kSZ studies \citep{Coulton_2024}. This apparent difference underscores that, while the exact nature and extent of baryonic feedback remain open questions, both kSZ and patchy screening are complementary and powerful tools for probing baryonic physics. To resolve these uncertainties, more precise observations coupled with improved theoretical models are essential.

As we enter an era of high-precision cosmology driven by forthcoming large-scale galaxy and cluster surveys such as Euclid \citep{EUCLID_2011}, Roman \citep{WFIRST_2011}, LSST \citep{LSST_2012}, and DESI \citep{DESI_2016}, accurate modelling of baryonic effects in large-scale structure (LSS) analyses will become increasingly important. The precision and accuracy in measuring the expansion rate and large-scale distribution of matter in the Universe will reach unprecedented levels. Moreover, data from surveys like eROSITA \citep[][X-ray]{eROSITA_2021}, Advanced ACT \citep[][tSZ, kSZ]{AdvACT_2016}, and the Simons Observatory \citep[][tSZ, kSZ]{Simons_2019} will offer tighter constraints on the baryon fraction of groups and clusters across different redshifts.

To fully harness the potential of these surveys for constraining cosmological parameters and exploring possible extensions to the standard model, it is crucial to ensure accurate modelling of LSS by rigorously accounting for baryonic effects through a comprehensive exploration of the ‘feedback landscape’. Recent studies, such as \cite{Bigwood_2024}, which performed a joint analysis of DES Y3 cosmic shear and X-ray baryon fraction constraints using \texttt{SP(k)}, not only improved cosmological constraints but also provided valuable insights into astrophysical feedback models. Moving forward, informative priors derived from observational constraints on the baryon fraction and other probes of hot gas will be vital in minimising the degradation of cosmological constraints and avoiding biases.

\section*{Acknowledgements}
The authors thank Alexandra Amon for helpful comments on a draft version of the paper. This work was supported by the Science and Technology Facilities Council (grant number ST/Y002733/1).
This project has received funding from the European Research Council (ERC) under the European Union’s Horizon 2020 research and innovation programme (grant agreement No 769130).
This work used the DiRAC@Durham facility managed by the Institute for Computational Cosmology on behalf of the STFC DiRAC HPC Facility (www.dirac.ac.uk). The equipment was funded by BEIS capital funding via STFC capital grants ST/K00042X/1, ST/P002293/1, ST/R002371/1 and ST/S002502/1, Durham University and STFC operations grant ST/R000832/1. DiRAC is part of the National e-Infrastructure.

\section*{Data Availability}

The data supporting the plots in this article are available upon reasonable request to the corresponding author. A Python implementation of the \texttt{SP(k)} model can be accessed at \href{https://github.com/jemme07/pyspk}{https://github.com/jemme07/pyspk}. 

The \texttt{SP(k)} model is also fully integrated into CosmoSIS \citep{zuntz_cosmosis:_2015} and can be provided for cosmological pipelines upon request. Those interested in using the ANTILLES simulations are encouraged to contact the corresponding author.




\bibliographystyle{mnras}
\bibliography{newbib} 





\bsp	
\label{lastpage}
\end{document}